\documentclass{INTERSPEECH2023}

\usepackage{hyperref}
\usepackage{algorithm}
\usepackage{algpseudocode}

\interspeechcameraready

\title{Conditional Online Learning for Keyword Spotting}
\name{Michel Meneses, Bruno Iwami}
\address{SiDi, Brazil}
\email{m.meneses@sidi.org.br, b.iwami@sidi.org.br}

\begin{document}

\maketitle
 
\begin{abstract}
Modern approaches for keyword spotting rely on training deep neural networks on large static datasets with \textit{i.i.d.} distributions. However, the resulting models tend to underperform when presented with changing data regimes in real-life applications. This work investigates a simple but effective online continual learning method that updates a keyword spotter on-device via SGD as new data becomes available. Contrary to previous research, this work focuses on learning the same KWS task, which covers most commercial applications. During experiments with dynamic audio streams in different scenarios, that method improves the performance of a pre-trained small-footprint model by 34\%. Moreover, experiments demonstrate that,  compared to a naive online learning implementation, conditional model updates based on its performance in a small hold-out set drawn from the training distribution mitigate catastrophic forgetting.
\end{abstract}

\section{Introduction}
\label{sec:intro}

Keyword spotting (KWS) concerns detecting a predefined set of spoken words in an audio stream \cite{szoke2005comparison}. Compared to general automatic speech recognition (ASR), KWS deals with a much more restricted predefined vocabulary. That explains why the industry has widely adopted KWS systems as the entry-point for more complex voice applications that target edge devices \cite{michaely2017keyword, yang2020multi, shrivastava2021optimize}, such as personal voice assistants (\textit{e.g.}, Bixby, Siri, Alexa). In those cases, the KWS system runs entirely at the edge in an always-listening mode. The goal is to detect commands that should trigger the rest of the application, which is usually more complex and, therefore, runs on another machine with considerably more computational power.

Modern approaches for KWS currently rely on training deep neural networks (DNN) to classify spectral representations of audio segments extracted from the input stream (\textit{e.g.}, log-Mel spectrograms, Mel-frequency cepstral coefficients) \cite{chen2014small, sainath15b_interspeech}. They typically depend on large training datasets of spoken keywords, either labeled or not, which should be as large and diverse as possible, so the training process can better cover the problem space \cite{meneses22_interspeech}. Therefore, while training, the DNN model is fed with thousands of positive samples representing each target keyword and negative samples representing other spoken words and environmental sounds that should not trigger the model.

Although that offline-training approach sounds reasonable, in real life, engineers must go through the arduous process of preparing the largest and most comprehensive dataset possible to better approximate the training distribution to the one expected in the final application scenario after the deployment of the model. Moreover, once in production, the model keeps unchanged for months after its deployment until its vendor schedules a periodic software update. During that meantime, changes in its input distribution, not anticipated during training (\textit{e.g.}, new prosodies, environment noises, and even microphone settings), may degrade its predictive performance.

This work argues that instead of attempting to anticipate eventual distribution shifts in production by arduously building large training datasets, an alternative is to update the pre-trained model on the fly at inference time as new data from the production environment becomes available. Indeed, the latest releases of popular frameworks for machine learning already support on-device training \footnote{\url{https://www.tensorflow.org/lite/examples/on_device_training/overview}}. The latest research on online-continual learning for keyword spotting \cite{xiao22_interspeech, huang2022progressive} focuses on adapting the model to new tasks different from the one it was pre-trained to solve. That additional complexity drives those works to elaborate online learning approaches that are arguably difficult to execute on tiny low-resource edge devices, such as online knowledge distillation and online multi-task learning.

Contrary to those references, this work restricts the online adaptation of a pre-trained keyword spotting model to the same KWS task, given a continual stream of speech generated by a distribution not learned by the model before its deployment. By performing that exploration, this work intends to directly contribute to the many keyword spotting systems widely available in the market, which must detect a fixed set of target keywords but underperform due to the non-stationary nature of the production environment. By restricting the scope of the problem, this paper also gains margin to explore a simple online learning algorithm based on conditional model updates via stochastic gradient descent (SGD), which is suitable for execution on low-resource devices. Experiments with audio streams in different scenarios illustrate the increasingly superior performance of the online learner over its frozen pre-trained base keyword spotter while mitigating the effects of catastrophic forgetting compared to its naive implementation.

The remaining of this document has the following structure: \autoref{sec:background} introduces the related literature on continual learning for keyword spotting, \autoref{sec:method} formulates the online learning algorithm for the same KWS task, \autoref{sec:experiments} details the experimentation methodology, \autoref{sec:results} presents and discusses the results of the experiments. Finally, \autoref{sec:conclusion} states the final remarks about this work.

\section{Background}
\label{sec:background}

In the previous decades, machine learning research has focused on offline methods that iterate throughout all the training samples available in a fixed dataset to induce a decision model. They typically assume those samples come from a stationary distribution, which would eventually generate the samples observed during the model consumption in production \cite{facelik2021}. As a result, the output models tend to underperform when presented with changing data regimes \cite{hadsell2020embracing}. Continual learning (also referred to as \textit{lifelong} learning) is a paradigm that attempts to emulate the human ability to incrementally learn as new data becomes available, hence addressing the non-stationary nature of the real world. When implemented in an online mode, that paradigm allows the learner to update itself during inference without interrupting the application \cite{liu2020learning}.

Some recent works have explored continual learning for keyword spotting. \cite{xiao22_interspeech} pre-trains a TC-ResNet model to learn 15 keywords from Google Speech Commands (GSC) dataset. Then, it explores a knowledge distillation-based strategy to update that model every time new task data becomes available. Those new tasks consist of new keywords from GSC the model has never seen. Its previous states work as teacher references for the current online training step, so the resulting model's performance on the previously learned tasks is not degraded. \cite{huang2022progressive} introduces a method that instantiates a sub-network for each new task incrementally presented to the model. Each sub-network should be capable of learning the individual target keyword defined by each task while sharing some knowledge to accelerate their learning process. That work follows the same experimentation methodology described by \cite{xiao22_interspeech}.

Similarly to what has been observed in recent research on general ASR \cite{yang22w_interspeech}, the related works on continual learning for keyword spotting focus on adapting the model to new tasks different from the one it was pre-trained to solve. More precisely, those new tasks refer to detecting new target keywords never seen by the model before its deployment. Hence, those works explore elaborated techniques for incrementally adapting the model to each new task without reducing its performance on the previously learned tasks. Such a problem is known in continual learning literature as catastrophic forgetting \cite{hadsell2020embracing}.

\section{Method}
\label{sec:method}

\subsection{Problem Statement}

This work assumes there is a model $M_{0}$ already trained on a labeled dataset $D = \{(X_n, y_n)\}^N_{n=1}$ of size $N$. Each pair $(X_n, y_n)$ contains the feature vector $X_n$ of a spoken word and its label $y_n\mid y_n \in \{\text{\textit{target word}}, \text{\textit{non-target word}}\}$. It assumes a binary-classification problem for simplicity and because that covers most of the keyword-spotting applications that serve as the wake-up systems of voice assistants. Nonetheless, one could easily extend it to multiple classes.

Once trained, $M_{0}$ is deployed to the target device and put into production. After its deployment, a set of new labeled samples $S = \{(X_i, y_i)\}^I_{i=1}$ becomes \textit{incrementally} available to $M_{0}$, \textit{i.e.}, it has access to each pair $(X_i, y_i)$ sequentially. Those new samples result from the input audio stream captured by the device's microphone. The goal is to incrementally update $M_{0}$ as those new samples continuously become available. It is also assumed that it would be impractical to update $M_{0}$ based on $D \cup S$ due to the large size of $N$.

Similar to \cite{yang22w_interspeech}, the problem statement assumes the online data stream is labeled. Naturally, that is not the case for most real applications. However, following that work, this one includes that same assumption to focus further discussions exclusively on the online learner design and to set an upper bound for its performance. In practice, one could implement simple strategies for estimating the labels of the incoming data as isolated modules \cite{jin2018unsupervised}.

\subsection{Conditional Online Learning}
This work explores a simple but effective approach for online continual-learning KWS via neural networks based on stochastic gradient descent (SGD) conditioned to the performance observed with a static hold-out dataset. Offline training methods for neural networks based on SGD usually iterate throughout a static dataset along multiple epochs via mini-batches. The motivation for processing the same data more than once is the limited number of samples available \cite{facelik2021}. In an online continual learning setting, incoming data is abundant. However, it is unreasonable to assume that data is independent and identically distributed (\textit{i.i.d}). As a result, directly updating the model via SGD based on that continuous data leads to poor convergence since the estimated gradients are biased \cite{zhao2018federated}.

To mitigate the effect of those biased estimates while taking advantage of that incoming data, the solution explored in this work considers a static hold-out dataset $D_v$ generated by the same \textit{i.i.d.} distribution of the training dataset $D$. Algorithm \ref{alg:proposed_method} describes that method, named COOL (an abreviation for \textbf{Co}nditional \textbf{O}nline \textbf{L}earning). Once both $M_0$ and $D_v$ have been deployed into the target device, during each step $i$ it receives the incoming pair $(X_i, y_i) \in S$ and update its target ($S_t$) and non-target ($S_n$) sample buffers accordingly. Once $S_t$ and $S_n$ have enough samples to compose a balanced batch $B$ of size $b$, the method updates the model's parameters via gradient descent optimization based on its instant loss for $B$. Before consolidating those new parameters, the method evaluates the model with the hold-out dataset $D_v$ and $B$. The new parameters become permanent only if there is a decrease in the loss on $B$ and $l_v'$, the loss on $D_v$, does not exceed $l_v$, \textit{i.e.}, the loss of $M_0$ on the hold-out dataset.

\begin{algorithm}[t]
	\caption{\label{alg:proposed_method}Conditional Online Learning (COOL)}
	\begin{algorithmic}[1]
		\Require $M_0$: pre-trained model
		\Require $D_v$: hold-out dataset
		\Require $S$: set of labeled samples from input stream
		\Require $b$: online batch size
		
		\State $j \gets 0$
		\State $S_{t}, S_{n} \gets \emptyset, \emptyset$
		\State $l_{v} \gets loss(D_v, M_0)$
		\ForAll{$(X_i, y_i) \in S$}
		
		\State Add $(X_i, y_i)$ into $S_{t}$ or $S_{n}$ depending on $y_i$
		
		\If{$S_{t}$ and $S_{n}$ have at least $b/2$ samples each}
		\State $B \gets$ the $b/2$ latest samples from $S_t$ and $S_n$ each
		\State $l \gets loss(B, M_{j-1})$
		\State $M_j = update(l, M_{j-1})$ \Comment{Via gradient decent}
		\State $l' \gets loss(D, M_{j})$
		\State $l_v' \gets loss(D_v, M_{j})$
		\If{$l_v' > l_{v}$ or $l' \geq l$}
		\State $M_j \gets M_{j-1}$ \Comment{Update not consolidated}
		\EndIf
		\State $j \gets j + 1$
		\State Discard the samples in $S_t$ and $S_n$
		\EndIf
		\EndFor
	\end{algorithmic}
\end{algorithm}

The continual online strategy described in algorithm \ref{alg:proposed_method} essentially reproduces the classical use of a hold-out set for regularizing the offline training of machine learning models. Nevertheless, from an engineering perspective, that strategy is simple enough to execute on edge devices with limited resources, given the modern frameworks for machine learning currently available. At the same time, its regularization aspect helps mitigate the undesired effects of distribution shifts compared to the naive SGD implementation (demonstrated in the next sessions), while demanding no hyper-parameter tunning besides defining the hold-out set $D_v$.

\section{Experiments}
\label{sec:experiments}

\subsection{Overview}
The experiments to evaluate COOL for keyword spotting have considered the following methodology:

\begin{enumerate}
	\item A small-footprint CNN model has trained on a KWS dataset of clean spoken words;
	\item The resulting pre-trained model has been evaluated on some noisy streaming test datasets;
	\item COOL has been evaluated with those same datasets on top of the pre-trained model as $M_0$.
\end{enumerate}

Both approaches had their accuracy on the test datasets registered for future analysis, along with the training and validation metrics of the pre-trained model. Additionally, COOL was compared to its naive online learning counterpart, which performs SGD updates without checking a hold-out set.

\subsection{Datasets}

\begin{figure}[t]
	\centering
	\centerline{\includegraphics[width=8.5cm]{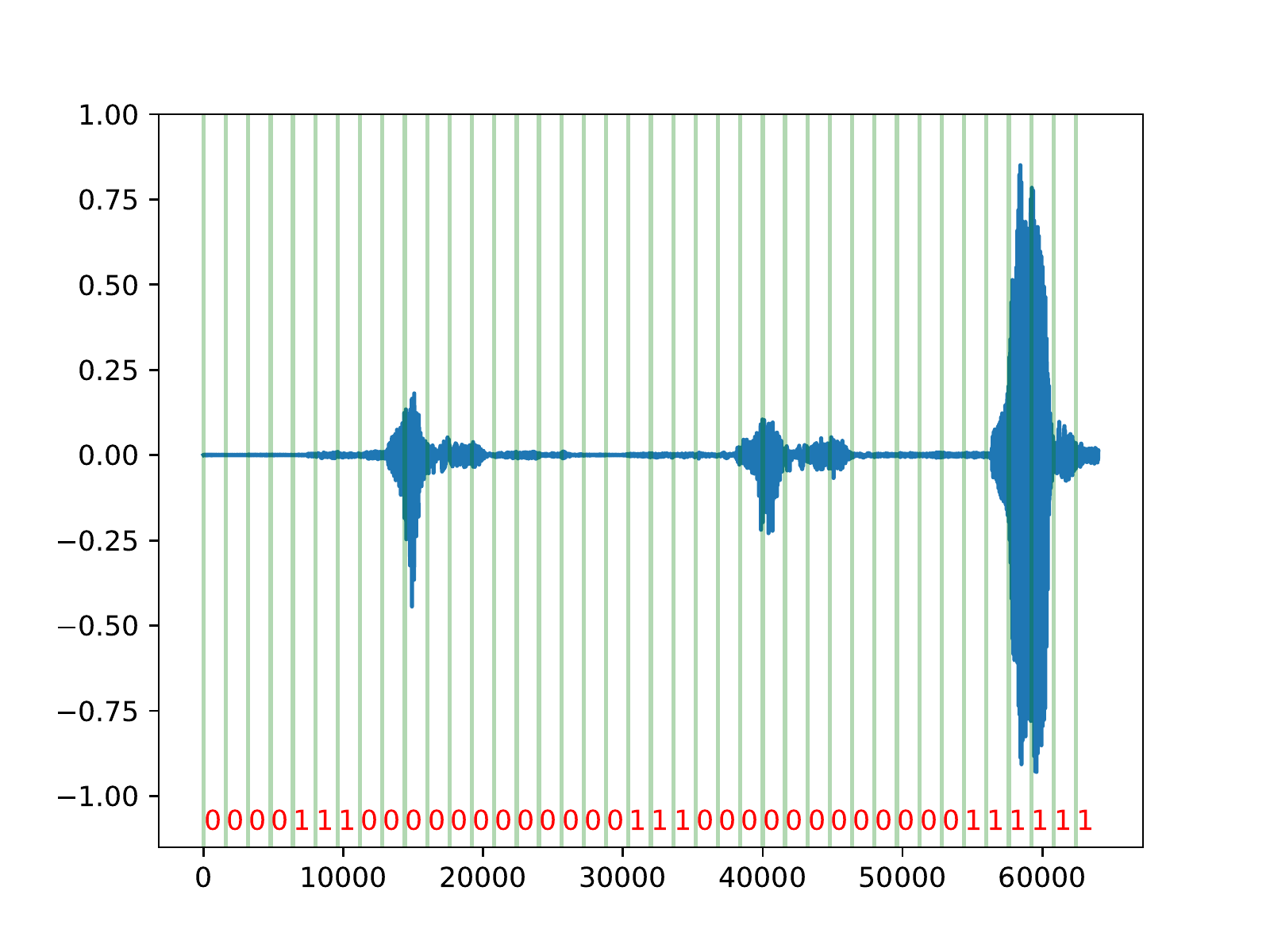}}
	\caption{Segment of a concatenated test audio file labeled according to the occurrence of the target word. Each label marks the first 100ms-duration frame of its respective input window of 1s.}
	\label{fig:test_db_labels}
\end{figure}

The experiments have considered Google Speech Commands v2 (GSC) \cite{warden2018speech} and DCASE \cite{DCASE2017challenge} datasets. GSC contains 105 thousand audio files of 35 different single spoken words. This work has considered a smaller version of it, which consists of 8 different words: "down", "go", "left", "no", "right", "stop", "up" and "yes". It has been used for training and testing the evaluated approaches. On the other hand, DCASE dataset has noisy audio files split into three categories: baby crying, gun shoot and glass break. It has been used as the noise source when mixing the clean test partition of GSC. Both GSC and DCASE contain audio files sampled at 16KHz.

Each of those eight words from GSC has originated a new binary classification task. Each task contains training, validation, and testing sets obtained by sampling the task's target word from the respective partitions in GSC. Both training and validation sets have the same words in clean conditions. Hence, the hold-out dataset used in algorithm \autoref{alg:proposed_method} for each task originates from that validation set. Each task randomly sampled their non-target words from GSC. As a result, there are eight tasks, each targeting one of those eight different words.

To simulate a data streaming, the final experiment considers a single long audio file per task, which is the concatenation of all their respective testing samples. All testing audio samples have been zero-padded with 500ms to avoid the cases where more than one word simultaneously occupies the model's input window. Each frame of those long files receives a label according to the presence of the target keyword. More precisely, a positive label indicates that at least 80\% of the target word occupies that frame (\autoref{fig:test_db_labels}). It has been mixed noise from DCASE with those test files at a SNR of 25dB.

\subsection{Settings}
The architecture of the base model $M_0$ is the \textit{cnn-one-fstride4} small-footprint convolutional network\cite{sainath15b_interspeech}. For each task, that model has its parameters updated on the training set for 20 epochs via Adam optimizer with a fixed learning rate of $1\mathrm{e}{-3}$, a batch size of 64 and a patient coefficient of 3 iterations for the early-stopping mechanism. It considers, as input features, 40-dim MFCC along 32 frames. The STFT frame length corresponds to 1K samples, and the frame step is 477 samples. The online batch size $b$ has been set to 16. It has been applied time-shift during training and validation.

\section{Results}
\label{sec:results}

\autoref{tab:train_loss_acc} presents the base model's final metrics after training for each target keyword task. It reached around $90\%$ validation accuracy for half of the tasks but only nearly $50\%$ for the other half. Nevertheless, the difference between training and validation metrics suggests the base model has not overfitted the training set of any task. \autoref{tab:test_acc} compares the average accuracy between COOL and its base model across the different test scenarios. For each of those, online learning has always initiated from the base model $M_0$, \textit{i.e.}, no knowledge is transferred between scenarios on this stage. As expected, COOL improves the average performance of the base model in all those scenarios. Moreover, it reduces its oscillation.

\begin{table}[t]
	\caption{\label{tab:train_loss_acc}Base model's final metrics after pre-training for each target keyword task.}
	\begin{tabular}{@{}ccccc@{}}
		\toprule
		\textbf{Target word}         & \textbf{Train loss} & \textbf{Val. loss} & \textbf{Train acc.} & \textbf{Val. acc.} \\ \midrule
		\textit{Down}       & $0,29$     & $0,31$    & $0,88$     & $0,89$    \\
		\textit{Go}         & $0,69$     & $0,69$    & $0,53$     & $0,50$    \\
		\textit{Left}       & $0,69$     & $0,69$    & $0,50$     & $0,56$    \\
		\textit{No}         & $0,69$     & $0,69$    & $0,50$     & $0,50$    \\
		\textit{Right}      & $0,20$     & $0,21$    & $0,91$     & $0,95$    \\
		\textit{Stop}       & $0,69$     & $0,69$    & $0,50$     & $0,50$    \\
		\textit{Up}         & $0,35$     & $0,34$    & $0,85$     & $0,89$    \\
		\textit{Yes}        & $0,19$     & $0,24$    & $0,92$     & $0,91$    \\ \bottomrule
	\end{tabular}
\end{table}

\begin{table}[t]
\caption{\label{tab:test_acc}Average accuracy comparison for all test keywords across the different test scenarios.}
\begin{tabular}{@{}lccc@{}}
	\toprule
	& \textbf{Base model}& \textbf{COOL}& \boldsymbol{$\Delta $}\textbf{\%}                      \\ \midrule
	Clean       & $0,52\pm0,27$ & $\mathbf{0,74\pm0,09}$& $30,13\%$ \\
	BabyCrying  & $0,53\pm0,28$ & $\mathbf{0,65\pm0,10}$& $18,29\%$ \\
	GlassBreak  & $0,53\pm0,28$ & $\mathbf{0,62\pm0,20}$& $15,21\%$ \\
	GunShot     & $0,53\pm0,28$ & $\mathbf{0,63\pm0,21}$& $16,20\%$ \\ \midrule
	Average     & $0,53\pm0,28$ & $\mathbf{0,66\pm0,15}$& $19,96\%$ \\ \bottomrule
\end{tabular}
\end{table}

\autoref{tab:test_acc_cumulative} shows COOL's average performance gain against its base model throughout a sequential test, \textit{i.e.}, all the audio tests for each keyword are concatenated into a single long audio file that encodes sequential environment transitions, hence there is knowledge transfering between scenarios. The first column of that table refers to the isolated performance gain observed for that particular scenario, whereas the second column presents the cumulative gain across all the experiments. Compared to the performance gain presented in \autoref{tab:test_acc}, the cumulative gain in \autoref{tab:test_acc_cumulative} demonstrates that the continual learning approach benefits from the knowledge obtained in previous scenarios. Moreover, it also demonstrates that the conditional update mechanism effectively mitigates catastrophic forgetting since the isolated gain observed when processing the clean scenario twice keeps stable.

\begin{table}[t]
	\caption{\label{tab:test_acc_cumulative}Average accuracy gain throughout the \textit{sequential} experiment.}
	\begin{tabular}{@{}lcc@{}}
		\toprule
		& \textbf{Isolated gain}   & \textbf{Cumulative gain} \\ \midrule
		Clean       & $28,85\%$       & $28,85\%$       \\
		BabyCrying  & $39,62\%$       & $33,96\%$       \\
		GlassBreak  & $43,40\%$       & $35,85\%$       \\
		GunShot     & $32,69\%$       & $35,85\%$       \\
		Clean       & $26,92\%$       & $34,62\%$       \\ \bottomrule
	\end{tabular}
\end{table}

\begin{figure}[t]
	\centering
	\centerline{\includegraphics[width=9cm]{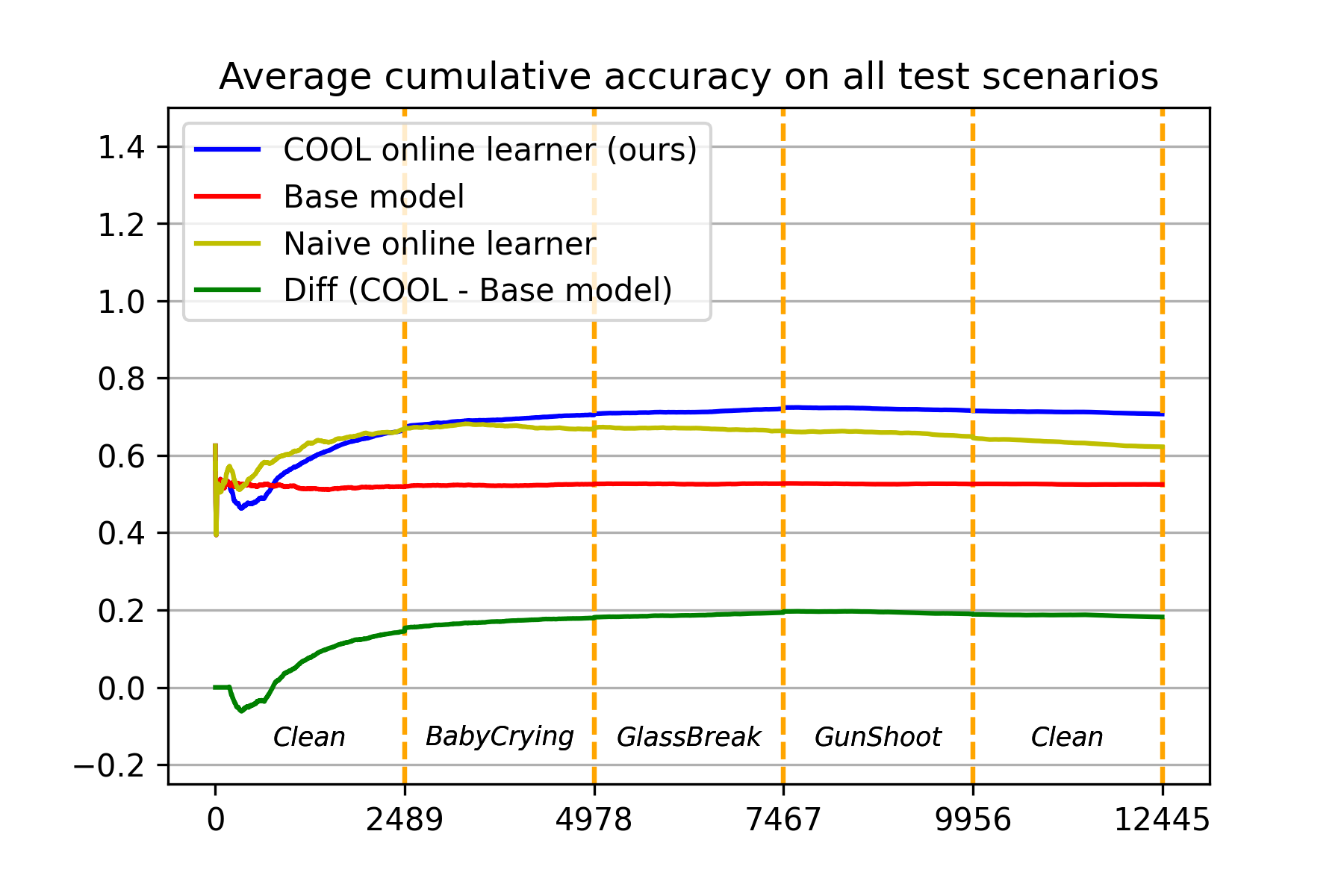}}
	\caption{Average cumulative accuracy curves throughout the sequential experiment. The y-axis indicates the cumulative accuracy while the x-axis refers to the test step. The vertical lines mark the transition between each scenario along the test audio file. The naive online leaner simply updates the base model without checking the loss with the hold-out dataset.}
	\label{fig:avg_cum_acc_all_kwd}
\end{figure}

Finally, \autoref{fig:avg_cum_acc_all_kwd} complements \autoref{tab:test_acc_cumulative} by illustrating the average cumulative accuracy gain observed during that long sequential experiment, including the performance of the naive online SGD implementation (\textit{i.e.}, without checking the loss with the hold-out validation set). Since it updates the base model after every inference step, that naive leaner initially overperforms COOL in the first scenario. However, its average cumulative accuracy starts decaying at the first change in the incoming data distribution since gradients estimated on that data are biased. On the other hand, by simply checking the loss of the updated model with the hold-out dataset, COOL mitigates the effect of those gradients while still taking advantage of the abundant incoming data to improve the accuracy of the KWS base model.

Overall, the previous results demonstrate that a simple approach such as COOL has the potential to significantly boost the performance of small-footprint keyword spotters, especially when presented with changing data regimes at inference time. Those results also suggest that simple regularization strategies effectively mitigate catastrophic forgetting. Naturally, limitations on this work's methodology affect those insights, such as the length of the testing audio stream and the assumption that the incoming audio samples are labeled. Even so, the results presented so far set an upper-bound performance gain that should encourage researchers and engineers to take a step further and build upon the method described in this work, especially since this is, to the best of our knowledge, the first work to set a practical baseline for online-continual learning in the context of streamed KWS with the same task targeting low-resource devices.

\section{Conclusion}
\label{sec:conclusion}

This work has evaluated a simple but effective online continual learning method for same-task keyword spotting based on SGD updates conditioned to the performance observed in a hold-out dataset. The experimental results demonstrate it dynamically improves the performance of a small-footprint pre-trained model on contrasting scenarios while receiving an input audio stream. Moreover, conditional updates improve the online learner's robustness against catastrophic forgetting. Due to its simplicity, that method can serve as a baseline for future research on online KWS and as an alternative for reducing false acceptance rates on commercial KWS applications. 

For future works, the authors aim to explore strategies for estimating the labels of the incoming data and evaluate how their uncertainty impacts the final performance of the online learner. The authors also intend to extend this study to the latest and more ellaborated online learning algorithms proposed in the related literature. Finally, the authors plan to implement those future steps in a real mobile application using the on-device training API recently available on TensorFlow Lite.


\bibliographystyle{IEEEtran}
\bibliography{refs}

\end{document}